# Evidence for 2-D π-bonded Surface Reconstructions on Si(111)

J. E. Demuth*

Naples, Florida 34114

Abstract: A silicene-like polymorph of the Si111 7x7 surface is proposed that resolves numerous experimental paradoxes and inconsistencies arising over the past 34 years. An analysis of the three established surface state charge densities from atom resolved spectroscopic imaging, including the 'forgotten' surface state at ~ -0.4 eV, shows features that are consistent with well studied 2-D silicenes. The bonding in this new structure as well as its physical nature are fundamentally different from the covalent surface bonding that is widely accepted. From its structural characteristics, this polymorph arises from significant $sp^2$-$sp^3$ hybridization in the top layer that creates a 'faulted' honeycomb motif of Si atoms. This top layer has an unusual periodic p-orbital structure that can inter-digitate with the terminal bulk 'dangling bonds' to create a 2-D π-bonded structure. This unusual bonding and new structure is important in understanding the nature of 2-D silicenes and 1-D honeycomb structures, as well as the conversion of the 1-D π-bonded chains of the 2x1 structure to the 2-D π-bonded 5x5 or 7x7 structures. Such extended π-bonded structures can lead to non-covalent dispersion interactions that have not been accurately included in previous semiconductor surface calculations.

Renewed interest in (111) surfaces has intensified with the recent discovery of a variety of hexagonal 2-D phases that are of potential technological importance.[1] Graphene[2] and silicene[3] are leading examples of 2-D materials whose hexagonal symmetry and planarity can lead to the formation of Dirac cones and Dirac fermions, i.e., mass less carriers, as well as other unusual properties.[4] Here evidence is presented for a new polymorph of Si111 7x7 whose structure is similar to certain silicenes and which has unusual π-bonding interactions with the surface atoms. Understanding the nature of this new family of 2-D silicene-like structures and the interactions that allow such bonding is thereby important.

One of the most fundamental aspects of quantum mechanics is utilized here to characterize the nature of this family of Si111 surfaces reconstructions. Namely, the atomic crystalline potential defines the stationary states that arise on the surface. As the Schrödinger equation prescribes, the electrons respond to the potential to find a configuration that minimizes the total system energy. The underlying symmetry of the wave-function and its spatial and energy structure provide a resulting probability distribution that can be measured through its energy and spatially dependent charge densities. Here, existing Scanning Tunneling Microscopy (STM) results are used to characterized the charge density distributions of the occupied 7x7 surface states, and are shown to be inconsistent with existing models.

In the mid 80's, the Dimer-Adatom-Stacking fault model (DAS) was introduced by Takayanagi[5] to account for the unusual 7x7 arrangement of Si atoms formed at the 111 surface. This now widely accepted structure was based on a Patterson and Fourier analysis of the diffraction intensities which indicated amongst other things a faulted structure.[5,6] From the various periodicities observed, a terminal layer was conceived using tetrahedral silicon atoms covalently bonded to minimized the number of dangling $sp^3$ bonds. Dimers were used to create a stacking fault and rebond the atoms in the top two layers. Such dimer rebonding produces large strain which is offset by strong electron pair bonds, i.e., covalent bonding.[7] The DAS model also describes a family of 2n+1 related structures called the DAS family with the same set of local atomic arrangements or building blocks.[8]

---

* jedemuth7x7@gmail.com

The DAS structure was found to be consistent with STM topographs,[9] and later with the observed atomic locations of two occupied surface states observed with STM Spectroscopy (STS).[9-11] Theory [12] and later full calculations within the local density approximation[13-16] confirmed the DAS structure and validated the DAS model. Due to its many successes, the DAS structure is widely accepted and describes many features of this surface.

As presented here, there is a third experimentally observed surface state that can now be understood based on the numerous past studies of the 7x7 surface. This third surface state was repeatedly observed for the 7x7 reconstruction: first, in temperature-dependent Photoelectron Spectroscopy (PES),[17], then reported in STS images,[10, 11] band mapped by angle resolved (AR)- PES,[18] seen in polarization depended AR- PES,[19] again found in STM measurements with an InAs tip [20] and later in STS measurements at T=78°K. [21] All calculations of the DAS structure confirmed the other two DAS surface states, but the third state was not. The third surface state has remained an enigma and is referred to here as the 'forgotten' surface state or FSS.

Several atom-resolved experimental measurements of the energy spectra above various atoms of the 7x7 show all three surface states,[21, 22] which as shown here, contradict the DAS calculations. Here, the energy, symmetry and atomic location of the FSS reveal the spatially delocalized nature of all three surface states. A new structure is proposed to explain this as well as other experimental findings based on the underlying features of a well characterized monolayer phase of silicene.[23] This new silicene-like structure provides a blueprint for the 7x7 structure and readily generates a new polymorphic family of reconstructions. The new structure is dominated by $\pi$-bonding interactions which make these surfaces a very different material than previously believed. Such differences have implications for the control, manipulation and even replication of surface structures not possible in a covalently bonded system.

The structure of this new polymorph is herein designated as the Digitated-Faulted-Adatom structure or DFA. The nature of the faulted surface arrangement of this new structure differs from the DAS structure and arises from the arrangement of sp2-sp3 hybridized atoms in just the top layer. This arrangement produces a different type of domain boundary separating the faulted and unfaulted sides of the surface unit cell and is responsible for an early paradox from x-ray standing wave measurements.[24] Namely, that there was no evidence of a stacking fault in the Si bilayer as proposed in the DAS model, despite clear evidence of a faulted structure in diffraction measurements.[5,6] The surface states and geometric arrangement of the DFA atoms suggest strong 2-D intra-layer interactions and surface bonding through an array of inter-digitated p-orbital charge densities, CDs, similar to $\pi$-bonding. Such bonding enables the top layer to be very flexible, readily form larger unit cell as well as accommodate surface stress.

Figure 1 shows a comparison of experimental spectral measurements of the surface DOS states, SDOS, for the 7x7 surface as determined by STS measurements of dI/dV normalized by I/V. The SDOS in (a) was obtained by averaging all the STS pixels over the unit cell when a tip of the highest resolution arose.[9-11] This spectra measured at 298 °K reveals all three surface states as marked. The lower panels (b) to (d) show atom resolved STS spectra obtained at 78° K [21] utilizing the same tip reformation process as used for the averaged SDOS above. The calculated SDOS [21] for the DAS model shown by the dashed lines.

The atom resolved STS results for the restatom and Adatoms[25] in the lower panel show marked differences between theory and experiment,[21] with the latter being nearly identical to earlier measurements[22] where comparable. In all DAS calculations[12-16] the FSS state is not found, nor is it evident in other calculations of the projected DOS of the surface atoms that include all the s- and p-states.[14, 16] The calculated restatom state [14, 16, 21] is well defined and only weakly coupled, if any, to the Adatom states. In contrast, the calculated SDOS of the dimers atoms [16, 18] (not shown) are strongly admixed into the Adatom states which is consistent with strong covalent bonding involving the 'dimers'.

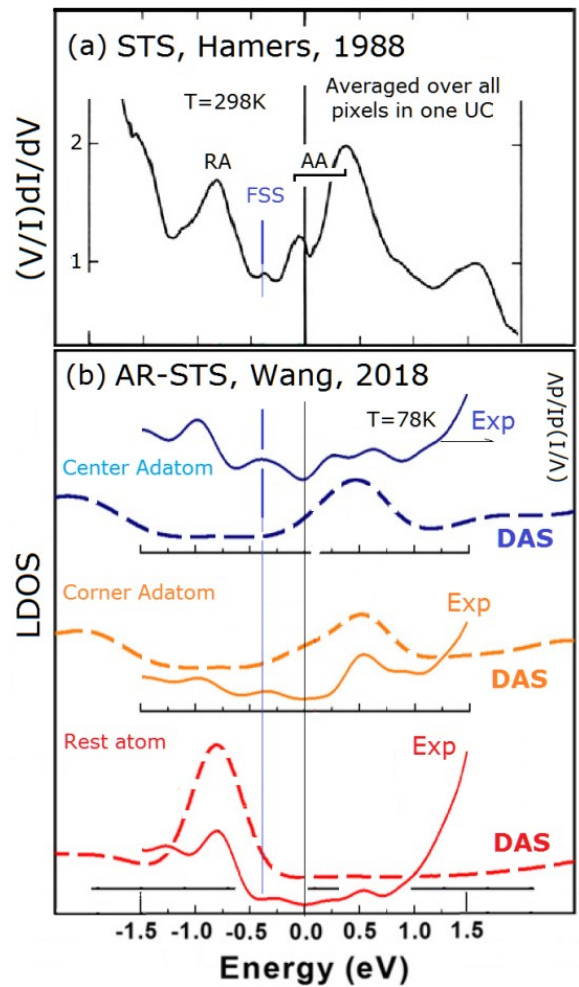

FIG. 1. (Color online) Comparison of the surface states of the 7x7 from (a) area averaged STS [11] and (b) DFT calculations and atom resolved STS. [21] In (a) the restatom (RA), the SFF and the Adatom (AA) manifold of states are indicated. A self energy correction of x1.6 is made to the calculated energy scale using the restatom as a reference versus the 1.4 value used earlier.[14] The theoretical spectra in

(b) have been further broadened from the spectra of earlier calculations[14, 16] by ~0.2eV to account for tip-derived broadening.[21]

In the atom resolved experiments,[21] the FSS is predominantly on the center Adatoms but also appears on the other atoms at slightly different energies. The restatom state at ~0.8 eV and Adatom state near $E_f$ are also admixed. These admixed states indicate that the basis states of the wave-function comprising these CDs, e.g., atomic orbitals or Slater determinants, occur on each of these atoms. Such admixed atomic charge densities will contribute to the delocalization of these states over its structure as is characteristic of the silicenes.[27-29] Depending on the degree of localization of these states, they may also reflect the top and bottom of various bands or sub-bands in angle resolved PES band mapping.[18, 19]

The spatial and energy characteristics of the three states also tell a similar story, namely, that the band structure of the DAS model differs from what is measured. A comparison is shown in Fig. 2 of an STS image for (a) the calculated DAS image [14] and (b) the experimental image [11] at an energies where the restatom has just appeared and the Adatom still shows a high intensity. Also shown besides each image is a schematic of how the Adatom and restatom CDs behave as a function of energy together with an inferred FSS in (b) that overlaps and appears between them. In (a) the FSS is not observed but instead the restatom and Adatom CDs appear to be separate and observed together over a narrow energy range of 0.15 eV centered at -0.44 eV [14]. Image subtraction of the calculated images at -0.44 and -0.58 eV or lower, confirms no additional CD between the restatom and Adatom.[26]

In the experimental images the Adatom state has merged with the restatom by -0.61 eV as shown in (c) and (d), with the Adatom state visible to -0.95 eV, a higher energy than calculated. The overlapping features at -0.75eV, and symmetry equivalent locations make the FSS look like a 'propeller' whereas in (a) at -0.73 eV only the restatom state is observed. Thus, the band structures of (a) and (b) must be different. In (c) and (d) a faint halo appears on the faulted side of the unit cell which is discussed later.

The FSS appears as if the Adatom CD is spatially dispersing to form the restatom state which has important significance as discussed later. Three different measurements [11, 14, 20] confirm this contrary behavior of experiment from that found in DAS calculations, with one [20] using energy filtered imaging clearly demonstrating that the FSS is not an artifact of either the tip or the STS procedures.

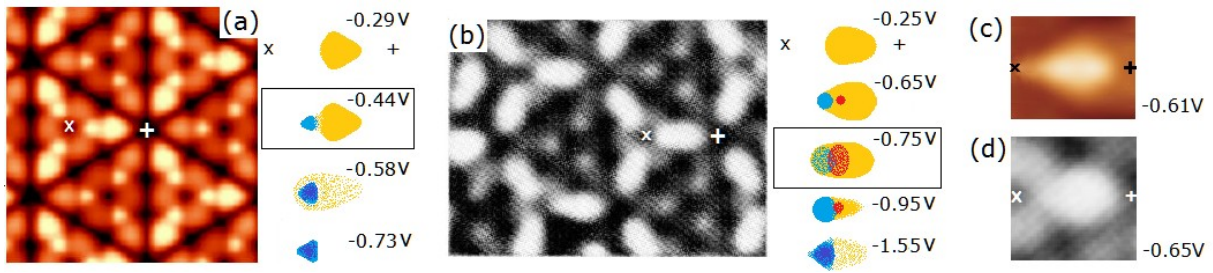

FIG. 2. (Color online) STS images of the surface state CDs as (a) calculated for -0.44 V [14] and (b) measured for -0.75 V.[11] X is the center of the unit cell and + is the corner hole, both for the 'faulted' side of the unit cell. Alongside each image is an enlarged schematic of the CDs between these points. Here the restatom CD is indicated on the left, the Adatom on the right and the FSS in between. The calculated energies shown are renormalized for self energy corrections as in Fig 1.

Panels (c) and (d) compare two experimental STS CDs [14, 11] averaged over equivalent directions to improve signal to noise.

The Adatom and restatom also show another distinctive property which may limit interpreting their shape, yet at the same time reflects another fundamental quantum mechanical property of electrons in solids. First, the calculated image shown in Fig. 2(a) considers s-wave tunneling and samples only the $\Gamma$ point of the Surface Brillion Zone, SBZ. As suggested by Sutter,[20] it is likely that part of the broadening or overlap of the CDs seen in (b) arises from sampling further into the SBZ due to the structure of the tip and its coupling to bulk derived block waves near the tunneling apex. However, given the flat bands found in AR-PES for the FSS, which are characteristic of a more localized state, any dispersion found must arise predominantly along the symmetry directions of nearby atoms in this 2-D lattice. Such energy dispersion with k, arises from the scattering and interference of waves by the atoms in a periodic potential.[30] The DAS model has no periodic atoms nor atom pairs to create or shift bond charge in the observed direction of this dispersion whereas the DFA structure does.

Figure 3 shows 5 vertical layers proposed here for the Si atoms in the DFA structure for the more compact 5x5 polymorph (which also shows a similar FSS.[31]) After an exhaustive search for possible modifications to the DAS model and earlier proposed models ended in failure, the DFA structure was developed from considering the well characterized 4x4 Si on Ag111 silicene system.[23] The DFA essentially starts from a faulted 3x3 honeycomb structure lattice matched to Si111, adding Adatoms similar to the dumbbell structures of free standing silicene[32] and finally extending this faulted honeycomb and Adatom arrangement to larger meshes. These features produce a DFA family of structures analogous to the DAS family. The fault in the DFA structure is simply the motif of this 2-D structure instead of the rebonding and distortions from the dimers of the DAS model. (Note that a related structure was also considered but excluded for lack of CD overlap. This was a DFA rotated 90 degrees that allows the rebonding of the atoms around the corner hole as in the DAS model.)

Both the DAS and DFA structures have an almost identical 2-D arrangement of Adatoms which would make them difficult to distinguish via STM topographs alone. As widely recognized, the strong tunneling signal from the charge densities of the Adatoms near $E_F$ dominate tunneling, [7, 13, 14] making it difficult to sense other atoms via STM. This domination of the STM image by the Adatoms also creates the charge density depressions along the domain wall that look like gaps arising from adjacent dimers in the DAS model.[26] However, one distinguishing feature is that the DFA Adatom has an unusual bond site with a 'pseudo' dumbbell atom in the silicon layer below. This likely accounts for the 0.15- 0.2A larger Adatom height above the bi-layer determined experimentally from that found in DAS calculations.[26]

In the DFA structure the atoms are aligned vertically, which is not the case in the DAS structure which also has strong lateral distortions due to rebonding to the dimers.[5, 7, 11-15] In addition, the DFA structure with its up and down pattern of the atoms, nestles well into the underlying 1x1 bulk Si bilayer. As shown in (c), this allows the inter-digitation of the sub-surface $sp^3$ 'dangling bonds' facing up with the periodic p-orbitals of the 2-D honeycomb that face down. These p-orbital interactions are herein defined as inter-digitated $\pi-$bonding or D-PB. $\pi$-bonding alone is estimated to be ~0.39 eV for hexasilabenzene [33] or ~0.5 eV as found for the dispersion bonding of benzene's $\pi$-bonds to Ag111.[34]

A reasonable possibility for the DFA is that the atoms around the corner hole drop down and compress slightly around the center of the corner hole ( by ~0.3 A) to improve their electronic interactions with the CD's below. This creates a 6 member 'cyclic" ring of rehybridized Si atoms that can better inter-digitate with the 6 upward facing Si111 sp3 'dangling bonds' directly below. This forms a more strongly bound cyclic D-PB 'sandwich' structure, defined here as 'silizene'. This conjugated structure essentially converts the native π- bonded chains of a honeycomb to an energetically favored cyclic structure.

This silizene compression also provides another important function in relieving the lateral stress in the honeycomb layer arising from the smaller buckling of the admixed sp2 top later. Meanwhile the atomic structure of the honeycombs in the center of the unit cell, leads to an weaker interaction since it sees half the number of the subsurface dangling bonds as in the silizene structure. Together these p-orbital interactions between the substrate and the 2-D honeycomb layer provide numerous π–bonds that hold the 2-D layer to the surface. In addition, the unusual bonding of the Adatom in the highly stable dumbell-like structure [32] likely contributes to stabilize and bond this 2-D silicene layer to the substrate.

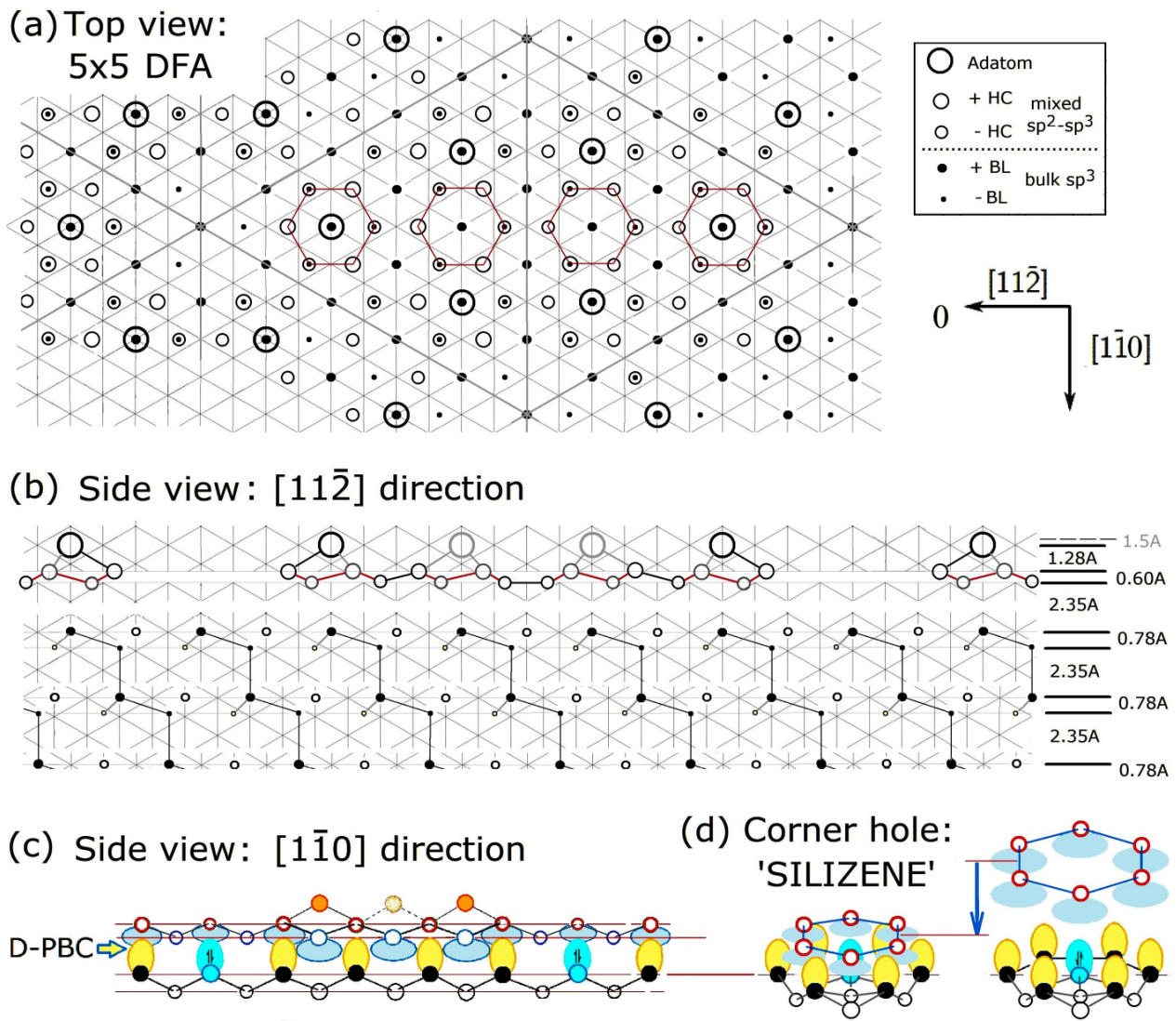

FIG. 3. (Color online) Structural model of the 5x5 DFA: (a) top view and side views (b) along the long diagonal and (c) the domain boundary. The relative sizes for the p-orbitals in (c) use the peak in the radial charge density of 2p hydrogenic wave-functions whose spatial extent is scaled to produce the correct bulk Si-Si bond distance. The hexagons shown in (a) reflect a few of the interior honeycombs. The spacings indicated in (b) are nominal bulk and silicene layer spacings with a 1.28 A Adatom height drawn. (d) shows a perspective view of the atoms comprising the corner hole with a small contraction in the bonds. The dangling bond in the corner hole likely has paired electrons as in the DAS model.

The overall structure and details of the DFA model such as the distortions of the silizene structure or the displacement of the substrate atoms to form the Adatom pseudo-dumbbell await improved structural analyses or total energy calculations that accurately include the dispersion forces in the D- PB structure.

The D-PB interactions that help bond the top layer to the surface atoms are analogous to the weaker interlayer bonding as occurs in graphite but with a different and likely stronger type of $\pi-$bonding, D-PB. It would not be surprising if this 2-D silicene layer could be peeled off with a 'sticky tape' that somehow substituted or altered the Adatom interaction with the substrate!

Fig. 4 shows a schematic of the CDs of the STM observed surface states for each model. Here, one also sees the local symmetries that produce the different domain walls of these structures.

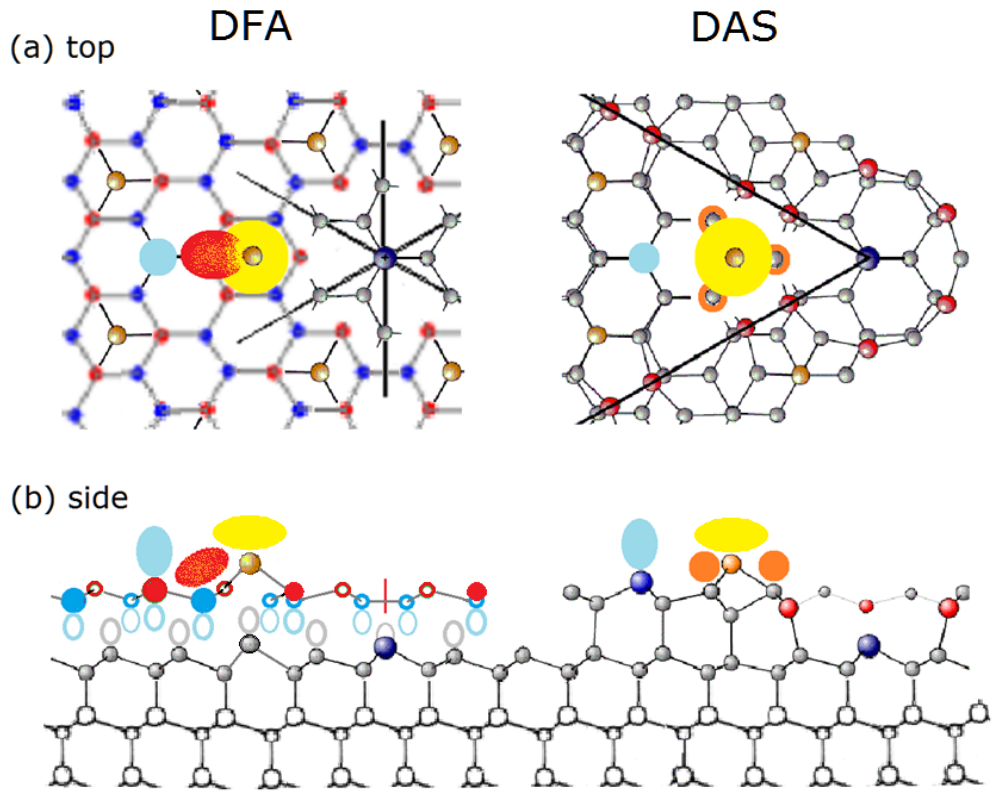

FIG. 4. (Color online) Proposed CDs on the faulted side of a unit cell for the DFA and DAS models. The elliptical lobes above the atoms in the DFA model represent p-like orbitals of the restatom, FSS and Adatom surface states. In the DAS model the smaller CDs below the Adatom correspond to a stacking fault state as projected[26] from LD calculations of the CDs of an intrinsic stacking fault[35]. The red (light) and blue (dark) atoms in the DFA honeycomb represent the higher and lower atoms respectively with an elevated substrate atom below the Adatom in (b) for the pseudo-dumbell. In (b) the smaller atoms represent atoms behind this cross section.

In the DFA structure the Adatom, restatom and FSS state are all involved to different degrees with in-plane bonding of the $sp^2$-$sp^3$ hybridized structure and share their electrons. Both the restatom and Adatom states appear in Fig.1 to be mostly localized at certain atomic sites but the FSS is not. Thus, while it is shown in Fig 4 to be a localized state, it is simply seen by STM at this location, either because of its spatially extension and/or its greater interaction and admixture with these Adatom and restatom states. Within a simple one electron model one can envision charge redistribution from the Adatoms to the restatom and now the FSS similar to that in the DAS model. Namely, the FSS fills last and at higher energies than the restatom due to larger Coulomb repulsion by the Adatom electrons.

The relatively weak PES signals of the FSS contrasts the stronger DOS features one usually sees in PES localized states or from typically flat bands (e.g., occurring near BZ boundaries). This low DOS for the FSS can be attributed to its being a crossing of the $\pi$- bands of the honeycomb layer that are now fully occupied. The AR- PES band dispersions of the 7x7 [19, 49] show many similarities to those of multilayer silicenes[50] in the same region of k-space. The later are proposed to arise from a $\pi-\pi*$ band crossing that starts at -0.25 eV. Performing an integration of the multilayer E versus k AR-PES spectra[50] about $\Gamma=0$ to as far as + - 0.4 A$^{-1}$ from $\Gamma=0$, produces a broad peak at ~0.4 eV that is almost identical to the FSS that is deconvolved from the Adatom and restatom PES peaks seen by Losio for the 7x7.[19] This suggests that the FSS state is delocalized laterally over the honeycomb, presumably from the $\pi$-bonds. However, whether these $\pi-$bands form a Dirac cone remains controversial.

The halo seen in Fig. 2b follows the expected bond charges for the DFA structure whereas the halo from the DAS calculation in Fig. 2a is larger and appears to arise from the Adatoms on each side of the unit cell. The energy filtered STM results[20] show the same size halo as Fig 2b as well as two weak FSS lobes observed around the restatom at slightly lower energies of ~0.6-0.7 eV below $E_f$. These additional lobes produce an underlying three fold symmetry about each restatom along the DFA bond directions. This is consistent with the FSS serving two roles: one as a split off state at the top of the bulk valence band ($\Gamma$-pt) at ~0.72 eV,[17-19] as found for the intrinsic stacking fault on Si111,[35] and another as the silicene intra-layer $\pi$-band. Interpreting the CDs for the empty states is even more precarious, given the complex structure and even more complex wave-function phases arising from such anti-bonding states.

One of many experimental results that supports the DFA model arises from an analysis of x-ray and electron diffraction data to refine the valence CDs of the 7x7 surface.[15] Namely, a bond-centered pseudo-atom (BCPA) formalism was used to refine the valence CDs from diffraction measurements. Figure 5 shows these results for (a) the DFT calculation which agrees with all prior DFT results, and (b) the BCPA refined CDs. Due to the large number of parameters in BCPA, various forms of averaging are required to reduce parameter space which may average out some of the 'refined' CD features.

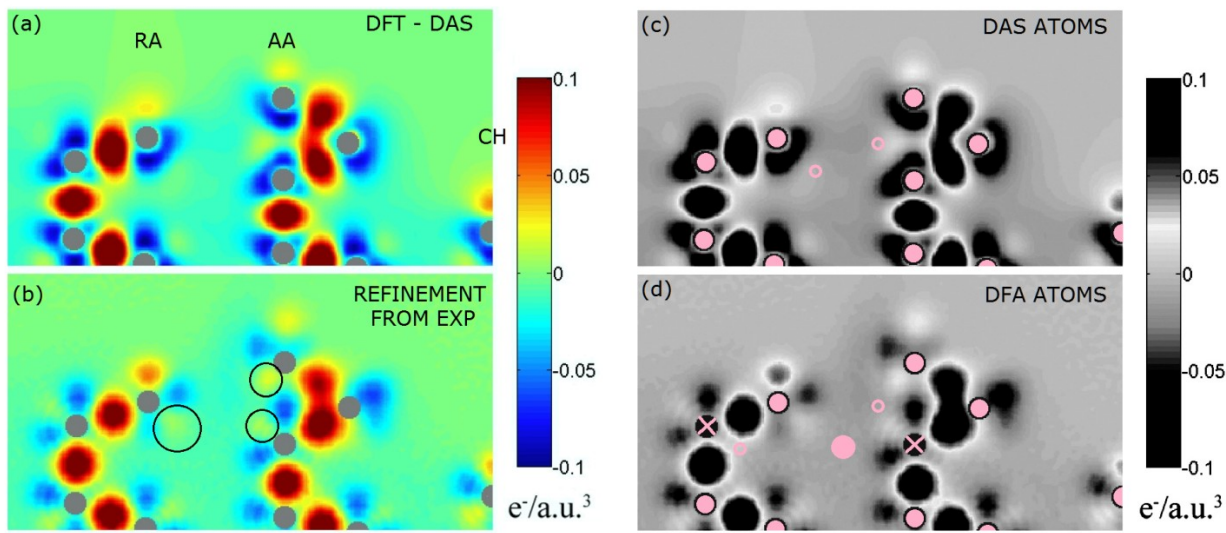

FIG. 5. (Color online.) Calculated CD differences for the restatom (RA), Adatom (AA) and corner hole atom (CH) on the faulted side of the 7x7 unit cell in (a) and (b) with (b) and (d) showing the experimentally refined CD differences from X-ray and electron scattering.[15] The atoms in (c) and (d) are a lighter tone while the open circles are the atoms behind this section. For the DFA structure in (d) the non-existent DAS atoms are crossed out and the new DFA atoms are shown.

Even with such averaging, the charge density refinement shows unexpected changes in the charge distribution from those found from the DFT calculations. The grey scale images on the left show the same area with the DAS atoms indicated in (c). In (d) the missing DAS atoms are crossed out and the DFA atoms added. The BCPA refinement is clearly a first order correction since the DAS structure is assumed in BCPA which frustrates a complete charge rearrangement. Never the less, this refinement as prescribed by the diffraction data, shows a redistribution of bond charge away from DAS bonds (c) to regions consistent with the intra-layer bonds of the DFA model (d).

In the DFA model the upward facing substrate $sp^3$ 'dangling bonds' shown in Fig. 3c and 4b can be likened to a 'spine' that can interact with pairs of down facing $p_z$ components of the honeycomb CDs, to form a new type of π- bond, the D-PB. This arrangement results in a regular array of substrate 'spines' along equivalent [1-10] directions that interact with the pairs of atoms along the domain walls as if the latter atoms were ribs. Stereo-chemically, these ribs have folded or zippered onto the periodic row of substrate dangling bonds with minimal stress to the lattice. The periodic substrate dangling bonds not only allow bonding to the substrate but also allow flexibility to expand to a large surface unit cell. Such folding is a very simplified analogue to the more complex folding in biological systems.

This new mode of interactions has direct implications to the details of how the 2x1 structure breaks its local symmetry starting at ~245°C,[36] and converts to the 5x5 by 280°C.[30] First, the symmetry directions of the 2x1 π- bonded chains (PBCs) and the D-PBCs are identical. In annealing the 2x1, an impurity or defect 'center' is proposed to initiate the ejection of a pair of atoms to form two Adatoms. This initiates the first domain boundary and a precursor to the corner hole. As this precursor expands its cyclic π- bonds, another pair of SI atoms are ejected and a second domain boundary forms via the zippering process. The production of Adatoms and zippering proceeds radially around the silizene precursor to eventually form the corner hole and the domain boundaries. When complete, it has created radial D-PBCs around silizene that eventually interlock to form a network of 2-D, D-PBCs that cover the surface.

There are many other experimental features that are problematical for the DAS model which the DFA model accounts for. A few include the geometric structure of the Adatom,[26] the reversible deconstruction of the Si111 7x7 with Si Adatom adsorption,[26] the CD symmetry of the empty states,[26] the structural features of several other meta-stable Si structures,[37, 38] the occurrence of [1-10] oriented PBCs for 1--D metals[39] on Si111 and alkalis on Si111,[40] as well as the unusual interaction potentials of He atoms[41] and positrons[42] in scattering from the 7x7 which are consistent with an $sp^2$ hybridized layer.[26]

Preliminary calculations of the 5x5 DFA structure using GGA VASP calculations with the PBE functional and parameters as defined earlier [43] show bonding interactions, but fail to converge to this layer stacking.[44] Systematic defects corresponding to ~1 atom per surface unit cell have been observed in the 5x5 and 7x7 structures [26] which may account for this. However, a more likely scenario is that the dispersion interactions of such a 2-D $\pi-$bonded system requires more accurate treatment than is possible with the exchange correlation functionals currently used in DFT calculations.[45,46]

For example, a theoretical study of the bonding of Benzene to Ag111 using a new method to correct for dispersion interactions [34] provides about a 0.5 eV larger bonding energy relative to PBE dispersion corrected DFT calculations. Given the magnitude of this interaction, more accurate dispersion corrected calculations could be a 'game changer' in predicting $\pi$-bonded silicon structures. Further, correlation effects may be more important on Si111 as the symmetry of this surface leads to large inter-atomic interactions whose contributions to the electronic energy can dominate the ground state .[47,48] Strong correlation effects together with longer range $\pi$-bonding interactions make this an even more challenging system !

The new structure proposed here represents a new paradigm for the interactions and bonding of the Si atoms in the outermost surface layers of Si111 where $\pi-$bonding can play a pivotal role in both 1-D or 2-D Si systems. Such bonding has important implications to the design of silicenes with low buckling and enhanced sp2 character, perhaps even to create a form of silicene 'sticky tape' noted earlier or to modify the charge transfer of the Adatoms to the FSS so as to produce massless carriers!

Acknowledgements: The IBM Watson Laboratory and ONR are gratefully recognized for their support during the author's Research career, and the University of Pittsburgh for access to their library system. Critical discussion with Ruud Tromp as well as comments from many colleagues about their work are acknowledged chronologically: O. Paz, J. M. Soler, L. D. Marks, J. Ciston, J. Ortega, R. M. Feenstra, N. Okabayashi, H. Ibach, M. Svec, B. Geisler, E. W. Plummer, I. K. Robinson, R.O. Jones, H. Tochihara, N. Takagi, G. Benedek, F. J. Himpsel, P. A. Bennett, Ph. Avouris, S. Y. Tong, W. A. Goddard, B. Voigtlander, P. Kocan, W. Jiang, P. Hansmann, E. G. C. P. van Loon, S. Meng and P. dePadova.

---------- ===== --------